\documentclass[twocolumn,prl,showpacs]{revtex4}
\usepackage{epsf}
\begin{document}
\title{Nematicity as a route to a magnetic field-induced spin density wave order; application to
the high temperature cuprates}
\author{Hae-Young Kee}
\email{hykee@physics.utoronto.ca}
\affiliation{Department of Physics, University of Toronto, Toronto,
Ontario M5S 1A7 Canada}
%\affiliation{Korea  Institute for Advanced Study, Seoul, Korea }
\author{Daniel Podolsky}
\affiliation{Department of Physics, University of Toronto, Toronto,
Ontario M5S 1A7 Canada} 
\date{\today}
\begin{abstract}
The electronic nematic order characterized by broken rotational symmetry
has been suggested to play an important role in the phase diagram of the high temperature cuprates.
We study the interplay between the electronic nematic order and 
a spin density wave order in the presence of a magnetic field.
We show that a cooperation of the nematicity and the magnetic field  induces a finite
coupling between the spin density wave and spin-triplet staggered flux orders.
As a consequence of such a coupling, the magnon gap decreases as the magnetic field increases,
and it eventually condenses beyond a critical magnetic field leading  to
a field-induced spin density wave order. Both commensurate and incommensurate orders
are studied, and the experimental implications of our findings are discussed. 
%separable.
\end{abstract}
\maketitle

{\it Introduction} ---
Typically, 
conduction electrons in strongly correlated materials are either localized due to interactions between them, 
or conduct with a uniform and isotropic distribution.
Can electrons in  metals assemble themselves and exhibit novel patterns seen in liquid crystals?
If so, what is the mechanism behind such a self-organizing pattern, and what are the effects of
such a metal on nearby phases? 
The electronic nematic phase characterized by an anisotropic conduction has been proposed 
as one such state. 
It was suggested that it plays a relevant role in determining
the superconducting transition temperature in the cuprates\cite{kivelson03RMP}, which motivated several studies on
the interplay between the nematic and d-wave superconducting (dSC) states.\cite{hykee-jpcm,eakim}  
An evidence of its existence in the cuprates was found by neutron scattering on YBa$_2$Cu$_3$O$_{6.45}$
(YBCO).\cite{keimer} 
On the other hand, the interplay between the nematic and the antiferromagnet, 
another nearby phase of the cuprates, has not been addressed.

In this paper, we study the relationship between the nematic and spin density wave (SDW) orders.
It is trivial to see that a direct coupling term between the two is not
allowed in the free energy, since they break distinctly different symmetries --
the nematic order breaks the rotational symmetry between $x$ and $y$ directions, while the AF order breaks
the translational, spin-rotational, and time-reversal symmetries.
Thus, a naive conclusion is that there is no strong influence between the two.
We show 
that the situation can be dramatically different when the time reversal symmetry
is broken by an external perturbation such as a magnetic field.
We find that there is a direct coupling between the SDW and spin-triplet staggered flux (tSF) orders
inside the nematic phase when the magnetic field is applied.
The tSF phase, like the (spin-singlet) staggered flux  (or sometimes called d-density wave) 
breaks the translational and rotational symmetries due to circulating currents with an alternating pattern. 
However, unlike the d-density wave state,
it does not break the time reversal symmetry, because up- and down-spin
circulations have opposite directions, which leads to the name of {\it spin-triplet} staggered flux.
A consequence of such a coupling is that the SDW order can be induced by the magnetic field via 
the nematicity.  This result applies to both commensurate and
incommensurate orders -- a finite coupling between incommensurate
spin density wave (ISDW)  and incommensurate triplet staggered flux (ItSF) orders occurs in the presence
of a magnetic field leading to ISDW order beyond a critical magnetic field.
We will discuss the experimental implications of our result in the context
of the high temperature cuprates.

\begin{figure}[htb]
\epsfxsize=8cm
\epsffile{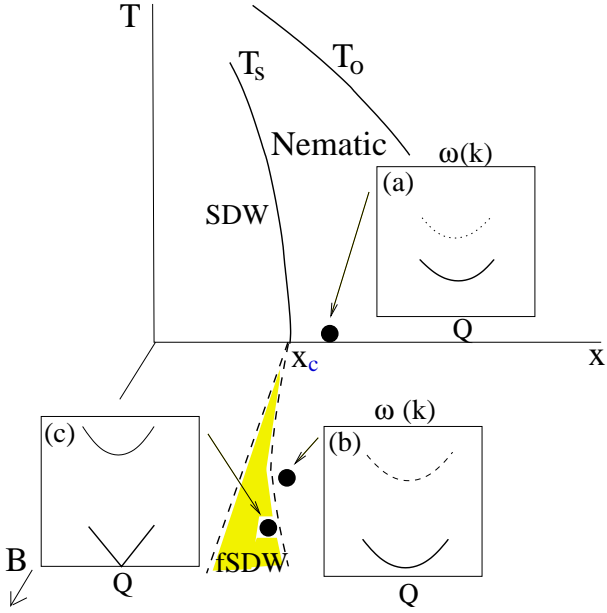}
\caption{Schematic phase diagram of SDW and nematic phases 
as a function of doping $x$, temperature, $T$ and magnetic field $B$.
The lines in the boxes are the dispersions of collective modes.
(a) In the nematic state,  both the SDW and tSF modes
are gapped.  (b) As the magnetic field increases, the magnon  gap decreases, while the tSF
gap gets pushed up.  Further magnetic field makes the magnon condense leading to
a field-induced SDW (fSDW) in the yellow region.
See the main text for further discussions on detection the collective modes for
incommensurate and commensurate SDW orders via neutron scattering 
and on the anisotropic intensity between $k_x$ and $k_y$ due to the nematicity.
\label{fig:phase-diagram}
}
\end{figure}

{\it Nematicity and commensurate SDW and tSF orders} ---
The nematic order, which breaks the $90$ rotational symmetry between
$x$- and $y$-directions, is defined as\cite{oganesyan,hykee,footnote}
\begin{equation}
{\cal N}_0= \frac{1}{2} \sum_{{\bf k} \sigma}  d({\bf k}) c^{\dagger}_{{\bf k} \sigma} c_{{\bf k} \sigma},
\end{equation}
where $d({\bf k}) = \cos{k_x} - \cos{k_y}$  (setting the lattice constant $a = 1$).
%parameter carries  
%charge 0, spin 0, and momentum 0, while AF order parameter carries charge 0, spin 1, and momentum ${\bf Q}$
%where $Q={\pi,\pi}$.
%
Inside the nematic state, marked by (a) in the phase diagram of Fig. \ref{fig:phase-diagram}, 
the quasiparticle Green's function is written as
\begin{equation}
G^{-1}({\bf k}, i\omega_n)= -i \omega_n + \epsilon_{\bf k}  -\mu,
\end{equation}
where  
\[
\epsilon_{\bf k}= -2 t (\cos{k_x}+\cos{k_y})  + 2t d({\bf k}) {\cal N}_0 -4 t^\prime \cos{k_x}\cos{k_y},
\]
%-2t'' (\cos{2 k_x}+\cos{2 k_y})$,
and $\mu$ is the chemical potential.
$t$ and $t^{\prime}$ represent the nearest neighbor and second nearest neighbor hoppings, respectively.
%$t$, $t^{\prime}$, and $t''$ represent the first, second, and third nearest neighbor hopings, respectively.

As shown in Fig.\ref{fig:phase-diagram}, 
the commensurate antiferromagnetic (AF) phase exists below a certain doping of $x_c$, and its order
parameter with a moment along ${\hat x}$ is given by
\begin{equation}
S^{i}_Q = \sum_{\bf k} c^{\dagger}_{{\bf k} \alpha} \sigma^i_{\alpha \beta} c_{{\bf k}+{\bf Q} \beta},
\end{equation}
where $\sigma$ is the Pauli matrix, $i = x,y,z$, and ${\bf Q} =(\pi,\pi)$.
%Note that the AF order parameter carries charge 0, spin 1, and momentum ${\bf Q}$.
%and breaks translation by a lattice constance, spin rotational, and time reversal symmetry. 
It is straightforward to check that there is no direct coupling between the nematic and AF
orders based on the distinctly different quantum numbers associated with their order parameters.

However,  when we consider the following tSF order parameter, the situation changes:
\begin{equation}
T^i_Q =  i\sum_{\bf k} d({\bf k})  c^{\dagger}_{{\bf k} \alpha} \sigma^{i}_{\alpha \beta}
c_{{\bf k}+{\bf Q} \beta}.
\end{equation}
The tSF order is characterized by circulating currents
with an alternating pattern like the d-density wave state, but up- and down-spins circulate 
in opposite directions.
Thus, there is no net charge current, but a finite spin current.
It breaks the translational, $x-y$ rotational, and spin rotational symmetries, but preserves
the time reversal symmetry.
This state was identified as a component of the 6 dimensional superspin which
can be rotated under the 15 generators forming the SO(6) group\cite{markiewicz} which includes
as a subset the SO(5) group suggested for a unified theory of the high Tc cuprates\cite{zhang}.
It was also discussed as one of non-zero angular momentum condensate states in Ref. \cite{nayak}.
%While the staggered flux is represented by a circulating charge current 
%on a plqutte with the same alternating pattern, in the tSF,
%there is no net charge current but a finite spin current
%due to opposite circulations of up- and down-spin.
In a dSC state,  both the tSF  and AF modes are gapped while
the AF gap, {\em i.e.} the magnon gap, is smaller than the tSF gap as shown in the box (a) in Fig. \ref{fig:phase-diagram}, due to the proximity of the AF phase in the phase diagram.

{\it Effects of a magnetic field in the nematic phase} ---
Since both ${\bf T}_Q$ and ${\bf S}_Q$ carry charge 0, spin 1, and momentum ${\bf Q}$, 
one may expect that there is a direct coupling between the two order parameters such as
\begin{equation}
{\cal F} = \gamma_{ij}({\cal N}_0, {\bf B}) S_Q^i T_Q^j,
\label{coupling}
\end{equation}
where the coupling $\gamma_{ij}$ is a function of ${\cal N}_0$ and the magnetic field ${\bf B}$
to respect the discrete symmetries of the system.
First,   the tSF order breaks $x-y$ symmetry, while the AF does not, thus the coupling
between them vanishes, except inside the nematic phase.
%\cite{footnote2}
Therefore, $\gamma({\cal N}_0=0, {\bf B}) =0$.
Second, the AF order breaks the time reversal symmetry, while the tSF order does not, which implies
$\gamma({\cal N}_0, {\bf B}=0) =0$.
Therefore, the coupling between the tSF and AF orders is finite only in the presence of
both a magnetic field and the nematic order, $\gamma({\cal N}_0 \neq 0, {\bf B} \neq 0) \neq 0$.
To linear order in ${\cal N}_0$ and ${\bf B}$, the form of coupling is found to be
$ {\cal F} \propto {\cal N}_0 {\bf B} \cdot ( {\bf S}_Q \times {\bf T}_Q)$, {\em i.e.} the index dependence of $\gamma_{ij}$ is given
by $\gamma_{ij}\propto \epsilon_{ijk}B^k$.
It is straightforward to check that such a term is allowed in the free energy based on 
the symmetry consideration discussed above. 

To compute $\gamma_{ij}$ in Eq. (\ref{coupling}), it is useful to introduce
$\psi^{\dagger}_{\bf k} = (c^{\dagger}_{{\bf k} \uparrow}, c^{\dagger}_{{\bf k}+{\bf Q} \downarrow})$.
In the basis of $\psi$, the Hamiltonian is written as
\begin{equation}
H^0_{nem} = \sum_{\bf k} \psi^{\dagger}_{\bf k}  \left[ (\tilde{\epsilon}_{\bf k} +B)\tau_3 -\mu_{\bf k} I \right]  \psi_{\bf k},
\end{equation}
where $\mu_{\bf k} = \mu - \frac{\epsilon_{\bf k} +\epsilon_{{\bf k}+{\bf Q}}}{2} = \mu_{{\bf k}+{\bf Q}}$,
$\tilde{\epsilon}_{\bf k} = \frac{\epsilon_{\bf k} -\epsilon_{{\bf k}+{\bf Q}}}{2}
= -\tilde{\epsilon}_{{\bf k}+{\bf Q}}$, and ${\bf B} = B {\hat z}$.
Note that the AF and tSF fluctuations couple to fermions as
\begin{eqnarray}
H^\prime= &=& g_1 \sum_{\bf k} S_{Q}^x c^{\dagger}_{{\bf k} \uparrow} c_{{\bf k}+{\bf Q} \downarrow} 
+ g_2 \sum_{\bf k} d({\bf k}) T_{Q}^y  c^{\dagger}_{{\bf k} \uparrow} c_{{\bf k}+{\bf Q} \downarrow} + h.c
\nonumber\\
&=& \sum_{\bf k} \psi^{\dagger}_{\bf k} (g_1 S_{Q}^x \tau_1 +g_2 d({\bf k}) T_{Q}^y  \tau_1) \psi_{\bf k},
\end{eqnarray}
where $g_1$ and $g_2$ are the interaction strengths in the AF and tSF channels, respectively.
We have chosen $x$- and $y$-component of the AF and tSF fluctuations assuming
that there is a small anisotropy such as Dzyaloshinskii-Moriya interaction aligning the staggered moment in the $x-y$ plane.
%\cite{yslee}

\begin{figure}[htb]
\epsfxsize=6cm
\epsffile{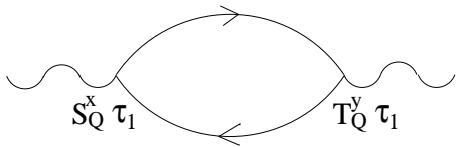}
\caption{Diagram to compute the coupling constant $\gamma_{xy}$ in Eq.~(\ref{coupling}).
\label{fig:nem-induced-order}
}
\end{figure}

Then $\gamma_{xy}$ in the nematic state and in the presence of a magnetic field is
obtained by computing the Feymann diagram shown in Fig. \ref{fig:nem-induced-order}.
\begin{equation}
\hskip -0.5cm \gamma_{xy} 
= g_1g_2 \sum_{\bf k} \frac{d({\bf k})}{\tilde{\epsilon}_{\bf k} + B} \left[ 
n_F(\tilde{\epsilon}_{\bf k} +B-\mu_{\bf k})- n_F(-\tilde{\epsilon}_{\bf k}-B-\mu_{\bf k}) \right],
\end{equation}
which is proportional to ${\cal N}_0 B$.
For example, for $t=1$, $t^\prime=-0.3$, and $\mu=-0.867$, 
$\gamma_{xy} = 2.36 {\cal N}_0 B g_1g_2$.
%Note that  $\gamma$ is finite only in the nematic state ${\cal N}_0$, in a non-zero field $B_z$.

%Then, the coupling between the antiferromagnetic and spin triplet staggered flux state  as shown
%in Fig. \ref{fig:nem-induced-order.eps} is 
%\begin{eqnarray}
%\gamma  &=& \sum_{{\bf k} i\omega_n}  d({\bf k}) {\rm Tr} 
%\left( \tau_1 G({\bf k} i\omega_n) \tau_1 G({\bf k} i\omega_n) \right)
%\nonumber\\
%&=& \sum_{\bf k} \frac{d({\bf k})}{2 \tilde{\epsilon}_{\bf k}} \left[ 
%n_F(\tilde{\epsilon}_{\bf k} -\mu_{\bf k})- n_F(-\tilde{\epsilon}_{\bf k} -\mu_{\bf k}) \right]
%\nonumber\\
%&=& 0
%\end{eqnarray}
%
%The coupling vanishes, because both the nematic and triplet staggered flux preserve time reversal
%symmetry, while the antiferromagnetic order does not. 
%Therefore, we expect that a external perturbation which breaks time reversal symmetry such as
%a magnetic field will induce a finite $\gamma$.
%%$d({\bf k})$ and $\tilde{\epsilon}_{\bf k}$
%changes sign under ${\bf k} \rightarrow {\bf k}+{\bf Q}$.

{\it Field-induced antiferromagnetism, collective modes, and neutron scattering} --- 
The free energy deep in the nematic state (static ${\cal N}_0$), expressed in terms of
the AF and tSF fluctuations, is given by
\begin{eqnarray}
{\cal F} &=&  	 \frac{1}{2m_s} 
	\left[(1+ {\cal N}_0) (\partial_x S_{Q})^2 + (1-{\cal N}_0)(\partial_y S_{Q})^2 
	\right] 
	\nonumber\\
&+&\frac{1}{2m_t} 
	\left[(1+ {\cal N}_0) (\partial_x T_{Q})^2 + (1-{\cal N}_0)(\partial_y T_{Q})^2 
	\right]\nonumber\\
&+&\frac{1}{2} \Delta_{s} S_{Q}^2 +\frac{1}{2} \Delta_{t} T_{Q}^2
  -\gamma S_Q T_Q,
\end{eqnarray}
where $\Delta_s$ and $\Delta_t$ represent the gap of AF and tSF, respectively, when a magnetic field is absent.

%{\it equation of motion--- effect of the coupling to the magnetic excitations}
%The equation of motion is given by
%\begin{equation}
%\partial_t \left(\frac{\partial L}{\partial (\partial_t S)}\right) +\nabla \cdot 
%\frac{\partial L}{\partial (\nabla S)} -\frac{\partial L}{\partial S} =0
%\end{equation}

%The  dispersions of collective modes are given by 
%\begin{equation}
%\omega = \sqrt{\frac{\omega_s+\omega_t}{2} \pm \frac{1}{2} \sqrt{ 
%(\omega_s -\omega_t)^2 + 4 \gamma^2}},
%\end{equation}
%where 
%\begin{eqnarray}
%\omega_s  &= & \Delta_s+ \alpha_s (1+ {\cal N}_0) k_x^2+ \alpha_s (1-{\cal N}_0) k_y^2,
%\nonumber\\
%\omega_t  &= & \Delta_t+ \alpha_t (1+ {\cal N}_0) k_x^2+ \alpha_t (1-{\cal N}_0) k_y^2.
%\end{eqnarray}
%
%The above dispersion implies that  the magnon gap decreases as the magnetic field increases, and
%it becomes gapless leading to a field induced AF state, when the following condition is satisfied:
%\begin{equation}
%|\gamma| = \sqrt{\Delta_s \Delta_t}.
%\end{equation}
%The leading contribution of ${\bf B}$ and ${\cal N}_0$ to $\gamma$ can be written as
%$\alpha(N_0,h) = \alpha_0 N_0 h$. 
%Assuming that $\Delta_s \propto (x-x_c)$, 
%the doping dependence of the critical field above which a field induced AF state occurs is given by
%\begin{equation}
%B_c \propto \frac{\sqrt{(x-x_c) \Delta_t}}{{\cal N}_0},
%\end{equation}
%which determines the phase boundary between the pure nematic phase and a field induced AF phase
%as shown in the dashed line in Fig. \ref{fig:phase-diagram}.

The above free energy becomes unstable to the formation of antiferromagnetism 
when the following condition is satisfied:
\begin{equation}
|\gamma| = \sqrt{\Delta_s \Delta_t}.
\end{equation}
%The leading contribution of ${\bf B}$ and ${\cal N}_0$ to $\gamma$ can be written as
%$\alpha(N_0,h) = \alpha_0 N_0 h$. 
Strictly speaking, the condensed mode is not a pure AF, but a mixture of AF and tSF.
However, if $\Delta_t \gg \Delta_s$, the mode is dominated by AF.
Assuming that $\Delta_s \propto (x-x_c)$, 
the doping dependence of the critical field for the onset of the field-induced AF state is given by
\begin{equation}
B_c \propto \frac{\sqrt{(x-x_c) \Delta_t}}{{\cal N}_0},
\end{equation}
which determines the phase boundary between the pure nematic phase and the field-induced AF phase
as shown in the dashed line in Fig. \ref{fig:phase-diagram}.

The collective mode dispersions are shown by different lines in the boxes in Fig.~\ref{fig:phase-diagram}.
The gapped magnon denoted by the thick line in the nematic state, as shown in Fig.~\ref{fig:phase-diagram}(a),
can be detected by neutron scattering, and its intensity should be anisotropic in momentum at a given frequency.
The anisotropic intensity of magnetic excitations was reported in YBCO, and 
interpreted as evidence of the nematic state.\cite{keimer,yjkao}
While the magnon can be detected by the neutron scattering, the tSF mode does not couple to
the magnetic field directly. Therefore, the tSF mode shown in the dotted line in Fig.~\ref{fig:phase-diagram}(a)
cannot be detected by the neutron scattering technique.  
However, in the presence of a field, the two modes couple as we have shown above.
The magnon gap is pushed down, while the tSF gap is pushed up, and the coupling also
generates a finite but small intensity of the tSF mode shown by the dashed line in Fig.~\ref{fig:phase-diagram}(b),
so the two modes should be in principle detectable by neutron scattering in a field.
A further magnetic field decreases the magnon gap as shown in Fig. 1(c) leading a field
induced AF order, while the tSF gap becomes larger.

{\it Incommensurate SDW and tSF orders} ---
Recently it was reported that the intensity of inelastic neutron scattering peaks at
the incommensurate wavevectors, ${\bf Q} = (\pi\pm\delta, \pi)$ in YBCO
increases when the magnetic field is applied.\cite{keimer2}
Here we show that the field induced SDW can be also applied to incommensurate orders for both collinear and
spiral cases.  We will discuss
how to distinguish spiral and collinear orders at the end of the section.

Let us first consider a collinear spin density wave order.
\begin{equation}
{\cal S}_{\bf Q}^x = \frac{1}{2}\sum_{\bf k} \left(c^{\dagger}_{{\bf k}-\frac{\bf Q}{2} \uparrow} c_{{\bf k}+\frac{\bf Q}{2}\downarrow} +
c^{\dagger}_{{\bf k}+\frac{\bf Q}{2} \uparrow} c_{{\bf k}-\frac{\bf Q}{2} \downarrow} + h.c \right)
\end{equation}
where we choose the spin density wave as ${\cal S}^x({\bf r}) \propto \cos({\bf Q} \cdot {\bf r})$.
Taking into account broken symmetries as before, the following incommensurate
tSF couples to ISDW linearly under the magnetic field.
\begin{equation}
{\cal T}^y_{\bf Q} = \frac{1}{2} \sum_{\bf k} {\tilde d}({\bf k},{\bf Q})
 \left( c^\dagger_{{\bf k}-\frac{\bf Q}{2}\uparrow} c_{{\bf k}+\frac{\bf Q}{2}\downarrow}
- c^\dagger_{{\bf k}+\frac{\bf Q}{2}\uparrow} c_{{\bf k}-\frac{\bf Q}{2}\downarrow} + h.c \right),
\end{equation}
where $ {\tilde d}({\bf k},{\bf Q}) = 2 (\sin{k_x} \sin{\frac{Q_y}{2}}
-\sin{k_y}\sin{\frac{Q_x}{2}})$.  We used a symmetric form of the order parameter for convenience. 
Note that the form factor of ${\tilde d}({\bf k},{\bf Q})$ 
has no longer a d-waveness when ${\bf Q}$ deviates from $(\pi,\pi)$
due to the constraint of the current conservation at each site.\cite{podolsky1} 

Note that due to the incommensubility $\delta$, a deviation from $\pi$, there is a finite coupling
between ISDW and ItSF under the field, even in the absence of the nematicity.
\begin{eqnarray}
\gamma_{xy} &=& g_1 g_2 \sum_{\bf k} \frac{{\tilde d}({\bf k},{\bf Q})}{2 (\tilde{\epsilon}_{{\bf k}-
\frac{{\bf Q}}{2}} + B)}
\left[ n_F (\tilde{\epsilon}_{{\bf k}-\frac{\bf Q}{2}} +B -\mu_{{\bf k}-\frac{\bf Q}{2}}) \right.\nonumber\\
&\,&\left.
- n_F (- \tilde{\epsilon}_{{\bf k}-\frac{\bf Q}{2}} -B -\mu_{{\bf k}-\frac{\bf Q}{2}})\right] 
\end{eqnarray}
%where $\tilde{\epsilon}_{\bf k} = \frac{\epsilon_{{\bf k}+{\bf Q}/2} - \epsilon_{{\bf k}-{\bf Q}/2}}{2}$
%and $\mu_{\bf k} = \mu - \frac{\epsilon_{{\bf k}+{\bf Q}/2} + \epsilon_{{\bf k}-{\bf Q}/2}}{2}$.
However, the coupling $\gamma_{xy}$ is proportional to $\delta^2$ which makes $\gamma_{xy}$ to be $0.022 B$
when $\delta = 0.12 \pi$.
On the other hand, when the nematicity is finite, $\gamma_{xy} = 2.4 {\cal N}_0 B g_1 g_2 $ similar to AF case.

Now let us study a non-collinear or spiral  spin density wave order. The spiral spin density wave
 can be written as
\begin{equation}
{\rm S}_{\bf Q} = \sum_{\bf k} \left(c^{\dagger}_{{\bf k}-\frac{\bf Q}{2} \uparrow} c_{{\bf k}+\frac{\bf Q}{2} \downarrow} +
c^{\dagger}_{{\bf k}+\frac{\bf Q}{2} \downarrow} c_{{\bf k}-\frac{\bf Q}{2}\uparrow} \right)
\end{equation}
where ${\bf S}_Q({\bf r}) = \cos{{\bf Q}\cdot {\bf r}} \hat{x} + \sin{{\bf Q} \cdot {\bf r}} \hat{y}$ with
spins lying in the $x-y$ plane. A similar spiral incommensurate staggered flux can be defined as
\begin{eqnarray}
{\rm T}_{\bf Q} =  \sum_{\bf k} {\tilde d}({\bf k},{\bf Q}) 
 \left( c^\dagger_{{\bf k}-\frac{\bf Q}{2}\uparrow} c_{{\bf k}+\frac{\bf Q}{2}\downarrow}
+ c^\dagger_{{\bf k}+\frac{\bf Q}{2}\downarrow} c_{{\bf k}-\frac{\bf Q}{2}\uparrow} \right),
\end{eqnarray}
Similar to the spiral spin density wave order, the spiral staggered flux can be viewed as  a pattern of
incommensurate staggering current where a spin quantization axis shifts from site to site determined by
${\bf Q}$.
Note that the coupling between the spiral spin density wave and the spiral staggered flux in
the presence of a magnetic field is the same as that for the collinear case.

Our results in general support a magnetic field-induced spin density wave including
both collinear and spiral orders.
Which one of these can be finally stabilized in the magnetic field requires a microscopic understanding
of cuprates which is beyond the scope of our current study.
However, we offer an experimental way to distinguish the collinear and spiral orders.
Note that the incommensurate collinear spin density wave couples to charge density wave order, which 
implies that there should be extra peaks at $(\pm 2\delta,0)$ or $(0, \pm 2 \delta)$
as the spin density wave order sets in at $(\pi\pm \delta,\pi)$ or $(\pi, \pi \pm \delta)$ under
the magnetic field.  
On the other hand, the spiral spin density wave does not induce a charge modulation.
Therefore, a further neutron scattering under a higher magnetic field would be a way to
determine the precise field-induced order.

{\it Other effects of the magnetic field} ---
The effects of the magnetic field on the order parameters deserves some discussion.
In the weak coupling approach, it was shown that the magnetic field enhances the nematic order for one
spin-component, but weakens the other component, and it shifts the phase boundary between
the nematic and isotropic metal.\cite{hykee2} However, deep inside the nematic state,
the field effect is negligible. 
%The effect of magnetic field on the nematic state in the strong coupling limit is not well known.
%
The magnetic field does not directly couple to the tSF order, because the tSF does not break
the time reversal symmetry. 
%and thus there is no linear field dependence.
Thus the magnetic field effect on the nematic and tSF orders does not affect our analysis.

The full effect of the direct interaction between magnetic field and AF order will be discussed later in the context
of high temperature cuprates, as it depends on circumstances.
%It also unifies the anisotropic magnetic excitations observed in YBCO by the neutron scattering.\cite{keimer}
However, the effect of the Zeeman coupling on the quasi-particles is rather straightforward to compute.
When the staggered moment lying in the $x-y$ plane, and the magnetic field is along
${\hat z}$, it changes the electronic dispersion as
$E_{\bf k} = \frac{\epsilon_{\bf k}+\epsilon_{{\bf k}+{\bf Q}}}{2}
\pm \frac{1}{2}\sqrt{(\epsilon_{\bf k}-\epsilon_{{\bf k}+{\bf Q}}\pm 2 B)^2 + 4 S_Q^2}$,
where there is no linear $B$ dependence. 
However,  when the staggered moment and the magnetic field are parallel, say along ${\hat z}$, then 
the dispersion is
$E_{\bf k} = \frac{\epsilon_{\bf k}+\epsilon_{{\bf k}+{\bf Q}}\pm 2 B }{2}
\pm \frac{1}{2}\sqrt{(\epsilon_{\bf k}-\epsilon_{{\bf k}+{\bf Q}})^2 + 4 S_Q^2}$,
with the linear $B$ dependence similar to the effect of the magnetic field on spin singlet condensate states 
 such as a charge density wave.

\begin{figure}[htb]
\epsfxsize=6cm
\epsffile{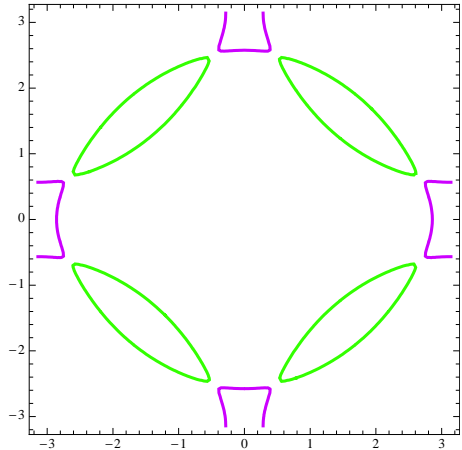}
\caption{The topology of a Fermi surface in a field-induced AF via the nematic order.
Note the anisotropy between the electron pockets near $(\pi,0)$ and $(0,\pi)$ due to
the nematic order. It is straightforward to generalize it for incommensurate SDW orders,
where the electron pocket remains the same\cite{footnote2}, while the hole pocket is sensitive to 
the incommensubility.
\label{fig:fermi-surface}
}
\end{figure}

{\it Fermi surface and Quantum oscillations} ---
The nematic phase in the absence of magnetic fields is metallic, and thus the field-induced SDW phase
is expected to be a metal.\cite{footnote3}
%, unless the induced order is so big and opens up a gap on the whole Fermi surface.
For example, as shown in Fig. \ref{fig:fermi-surface}, 
in the field-induced AF coexisting with nematic order, there are elongated pockets
--- one electron-like and one hole-like. It is straightfoward to generalize the Fermi surface
for a sprial order, where the electron pocket is qualitatively the same.

In a metal with a closed Fermi surface, 
magnetization $M$  and conductivity oscillate in $1/B$ as
\begin{equation}
M \propto \sum_{n} {\it A}_n \cos{\left[\frac{n ({\cal A}_{\epsilon_F}+\phi)}{B}\right]},
\end{equation}
where ${\cal A}_{\epsilon_F}$ is the area of a closed Fermi surface at the chemical potential.
Since there are two types of pocket, the primary periodicities ($n=1$) 
are determined by the size of both the hole and electron pockets.
Using the same parameter set of $t=1$, $t^\prime =-0.3$, $\mu=-0.867$
for $\gamma$, and setting $\langle S_Q^x \rangle = 0.07$ and ${\cal N}_0=0.05$, 
we obtain the periodicity of $540T$
and $900T$ for the electron and hole-pocket, respectively, as a unit cell of square lattice
$ \sim 3.82 \times 3.89 A^2$.
Due to the nematic order, the electron pockets around $(\pi,0)$ and $(0,\pi)$ are
elongated along different directions, but their area is the same yielding {\it a single}
frequency of the quantum oscillations.
As discussed above, in a Fermi liquid,
the Zeeman effect generates two different phases $\phi_\uparrow$ and $\phi_\downarrow$ (up and down spins
have different phases), due to the linear $B$ dependence in the electronic dispersion.
However, in the field-induced AF with a staggered moment lying in the $x-y$ plane, 
there is no linear $B$ dependence, implying a single phase of oscillation.

It is tempting to argue that our finding is relevant to
the quantum oscillations observed in high temperature  YBa$_2$Cu$_3$O$_{6.5}$ 
and YBa$_2$Cu$_4$O$_8$.\cite{taillefer1,taillefer2,proust,jaudet,bangura,sebastian}
Recent quantum oscillations reported in YBCO around $10 -12 \%$ doping were striking and
contrast with the Fermi arc - a truncated Fermi surface - reported by
the angle resolved photoemission spectroscopy.\cite{damascelli,shen}
There have been a few theoretical proposals which attempt to explain the observed quantum oscillations, including 
the d-density wave as a pseudogap state\cite{chakravarty}, the staggered flux 
in a vortex liquid\cite{palee}, stripe formation
\cite{norman}, and a field-induced antiferromagnet\cite{fuchun}.

%For the AF order with the staggered moment lying in  $x-y$ plane, 
%the understanding of $z$-axis field effects requires 
%a study of the competition between the dSC and AF, another important ingredient of physics.
As shown in Ref. \cite{fuchun},  the competition between the dSC and AF orders
also leads to a field-induced AF phase, 
where  the coexistence of AF and dSC states between $ 0 < x \leq 0.1$ is important to induce
a staggered moment beyond $10\%$ doping.
In our study, the AF and dSC coexistence is not a necessary ingredient in achieving
a field-induced AF order. Nonetheless, the competition would
help to lower the critical value of the magnetic field at which the AF order sets in.
While the mechanisms of the field-induced AF are not shared by the two studies,
both results favor a field-induced AF order.

%Here we show that the field induced antiferromagnet via nematicity is a strong candidate
%for the origin of both quantum oscillation and anisotropic magnetic excitations.
%Fermi arcs in underdoped cuprates, which do not seem to be consistent with
%the observed quantum oscillations. However, one should note that ARPES measurments
%were done in zero field and above $T_c$, while the quantum oscillations were observed
%in high fields and extremely low temperatures. 
%In the absence of filed, the ground state is  superconducting around $10 \%$ doping,

Among the commensurate orders, it is worthwhile to note that both the d-density wave (spin singlet staggered flux)
and AF orders lead to a qualitatively
similar Fermi surface topology originated from the  broken translational symmetry.
As proposed in Ref. \cite{podolsky}, 
the ortho-II potential can be used to test two different scenarios, the d-density wave order
versus AF order.
It was reported that both ortho-II-free, YBa$_2$Cu$_4$O$_8$ and ortho-II, YBa$_2$Cu$_3$O$_{6.5}$ 
show a primary oscillation with  
a periodicity of $660T$ and $540T$, which supports that the primary oscillation is from a pocket
insensitive to the ortho-II potential. Within our picture, the electron pocket is not
affected by the ortho-II potential. On the other hand, the hole pockets are modified by
the ortho-II potential as shown in Ref. \onlinecite{podolsky}, and the topologies of the modified hole pocket 
are different for the AF and d-density wave orders.  
Thus, a detailed study of the magnetic breakdown due to the ortho-II potential would
be another way to differentiate a field-induced AF via the nematicity
and the d-density wave order.\cite{jm}

In summary, 
we show that nematicity is a route to achieve a field-induced spin density wave order in the presence of
a magnetic field. An induced spin density wave is obtained via the coupling between the spin density wave mode and
the spin triplet staggered flux mode which are both gapped in the nematic phase.
Such a direct coupling between the two orders occurs only when a magnetic field is applied,
and the coupling strength increases linearly to the magnetic field. 
A further increase of the magnetic field eventually 
condenses the magnon leading to a field-induced spin density wave order, while the gap of the triplet staggered flux 
mode becomes larger.
A neutron scattering study under a high magnetic field will offer a precise order among
possible commensurate and incommensurate orders discussed above.
Our study indicates that the nematicity plays an important role in understanding
the phenomena observed in the high temperature cuprates, in particular in the context
of anisotropic magnetic excitations and quantum oscillations observed in YBCO materials.

This work is supported by
NSERC of Canada, Canadian Institute for Advanced Research, and Canada Research Chair.


\begin{thebibliography}{100}
\bibitem{footnote} While the electronic nematic order refers to a spontaneous broken symmetry, our current results
can be also applied for a nematicity arising from an external perturbation such as orthorhombicity.
\bibitem{footnote3} When the induced AF gap is large, it is possible to open a gap on a whole Fermi surface. However,
such a large gap would require an enormous magnetic field.
\bibitem{footnote2} For the collinear case,  a higher order mixing of bands is ignored,
assuming that magnetic breakdown occurs for such a small band gap.
\end{thebibliography}
\end{document}